\documentclass[12pt]{article}
\usepackage{epsfig, graphicx}

\begin{document}
\title{Domino Magnification}
\author{J. M. J. van Leeuwen\\
Instituut--Lorentz, Leiden University,  P. O. Box 9506, \\
2300 RA  Leiden, The Netherlands} 
\date{\today}
\maketitle

\begin{abstract}
The conditions are investigated under which a row of increasing dominoes is able to 
keep tumbling over. The analysis is restricted to the simplest case of frictionless dominoes 
that only can topple not slide. The model is scale invariant, i.e. dominoes and distances 
grow in size at a fixed rate, while keeping the aspect ratios of the dominoes constant. 
The maximal magnification factor for which a domino effect exist is determined as a function of the
mutual separation. 
\end{abstract}

\section{Introduction}

The domino effect has received considerable attention because it is spectacular and
seems to be governed by simple physics \cite{banks, efthimiou, walker}. However,
careful experiments \cite{mclachlan,stronge, larham} are not so easy. Measuring
the propagation speed of the order of a meter per second of a train of falling dominoes,
is already too difficult to observe by elementary means.
Equally, the theory of toppling dominoes is not as simple as the phenomenon looks like
and cannot be reduced to the straightforward application of the conservation laws of 
energy and angular momentum.
 
The discussion of a row of falling  dominoes involves two aspects: the tumbling motion 
and the collisions. The first to realize that the domino effect is a collective phenomenon was 
D.~E.~Shaw \cite{shaw}.  He noted that the dominoes lean on each other after collision. 
Consequently the collision of dominoes is fully inelastic and this is the main source of
energy losses during the process. It is tempting to invoke conservation of angular 
momentum at the collision as in \cite{shaw}, but this does not apply, since colliding dominoes
topple around different axes \cite{jmjvanl,vanleeuwen}.

The reason to come back on the domino effect is a question posed at the Dutch
Science Quiz 2012 reading: ``{\it how many dominoes does one need to topple a domino 
as tall as the Domtoren?}'' (a tower of 112 meter high). A restriction was that the dominoes 
have the same aspect ratios as that of standard dominoes. This restriction is essential for
an unique answer, since the thinner the domino (with fixed height) the easier it topples.
The idea behind the question is that every successive domino is a factor $r$ larger than the 
preceding one. The dominoes grow in size as a geometric series $r^n$ and the correct answer 
is the $n$ that makes $r^n$ times the height of a standard domino equal to 112 meter.
Whitehead \cite{whitehead} describes a demonstration with a magnification factor $r=1.5$.

In order to turn this problem into a scientific problem, one has to make further specifications.
One is a rule for the distance between successive dominoes. Although not specified in the 
problem, the logical choice is to scale the distance with the size. To let the distance grow 
with the size of the domino is reasonable, the larger the domino, the more space one needs 
between them. Moreover, there will be an optimal distance in each cycle and this optimum 
will scale with the size of the dominoes. So we restrict ourselves to a scale invariant model.
 
In order to make the mechanical model precise, a few more conditions have to be imposed:
\begin{enumerate}
\item The dominoes may not slide with respect to the ground. It is clear that sliding 
is an energy loss that impedes the domino effect. Infinite friction with the floor
is favorable and can basically be reached in practice. Thus the only possible motion
is to topple over. So the state of domino is given by a single angle $\theta$ measuring the
deviation from the normal to the ground. 
\item The dominoes stay in contact with each other after the collision \cite{shaw}. This makes the 
collisions fully inelastic. In practice this condition is fulfilled as the movie of the 
the demonstration with wooden dominoes shows \cite{jekel}. Inelastic collisions are the
main energy drain. Steel dominoes would do better, but are too costly to produce in large sizes. 
We take this condition as a constraint on the maximum possible $r$.
\item After hitting a new domino, the dominoes slide frictionless over each other. 
It is of course not realized with wooden dominoes, but it would be not difficult to
smooth the surfaces and minimize the friction by lubrication. Inclusion of friction
is quite well possible \cite{vanleeuwen, jmjvanlc}, leading to a substantial complication in the
calculation, while adding little to the understanding of the mechanism.
\end{enumerate}

We start the discussion with the elementary process of one domino colliding with 
another at distance $s$ and tumbling together downwards. This provides the setting for
the longer train of falling dominoes. The dimensions of the dominoes are denoted by
height $h$, the thickness $d$ and  the width $w$. We will consistently work with 
dimensionless distances to ease the calculations.  The parameter $w/h$ hardly enters
in the calculation. The width is only co-determinant for the mass $m_i$ of the dominoes. 
As in all gravitational phenomena, the total mass of the object drops out of the equations, 
but not the mass distribution. We consider two (extreme) cases: $q=4$, referring to 
massive dominoes where the mass is proportional to the volume and $q=3$, referring
to hollow dominoes, with a mass proportional to the surface. The latter case seems curious, 
because standard dominoes are massive. However in an attempt to topple a very large domino 
\cite{jekel}, the larger dominoes have to be hollow for practical reasons.

As mentioned the dominoes and their mutual distance all have the same aspect ratios: 
$w/h, s/h$ and $d/h=0.14583$, which is taken equal to the ratio of standard dominoes.
Each new domino is $r$ times larger than 
its predecessor. One could consider varying $s$ and $r$; it would make the discussion
considerably more involved and not so much richer, since we investigate the largest
possible magnification factor and it turns out that this occurs at a well defined value of $s$.
So deviating from the optimal $s$ and $r$ makes the domino effect less effective. 
Restricting ourselves to the scale invariant case allows to discuss only one cycle in the
domino effect, all the others are similar.

A dimensionless time is somewhat more delicate. The natural time unit reads
\begin{equation} \label{a1}
\tau = \sqrt{I/mgh} = \sqrt{(h^2+d^2)/(3 g h)},
\end{equation} 
where $I$ is the moment of inertia of the foremost domino. This is the appropriate
time scale for one cycle, the next is a factor $\sqrt{r}$ larger, the previous a factor 
$\sqrt{r}$ smaller. In the formulas we will not express time in terms of the unit $\tau$,
as this complicates the connection between different cycles.

The description of the train of falling dominoes centers around the tilt angle $\theta$ 
of the foremost domino as function of the time $t$. It starts at $\theta(0)=0$ and ends
at $t_f$ at collision angle $\theta(t_f)=\theta_c$, where the foremost domino looses its 
role as head of the falling train of dominoes. $\theta_c$ is given by
\begin{equation} \label{a2}
\sin \theta_c = \frac{r \, s}{h}.
\end{equation} 
After $\theta_c$, the domino continues to fall, but it becomes a slave, just as the previous
domino was of the foremost. We number the dominoes with respect to the foremost 
falling domino, which gets the number 0, or by default no number. So $\theta=\theta_0$
is the tilt angle of the foremost domino, $\theta_1$ that of its predecessor and so on. 
The calculation of $t_f$ gives the propagation speed of the domino effect. It is of no concern
here. We only note that $t_f$ will be proportional to $\tau$ defined in (\ref{a1}) and therefore
$t_f$ will increase with a factor $\sqrt{r}$ in every cycle. This slowing down of each cycle 
is the beauty of the record attempt \cite{jekel}, where one can see {\it ad oculos} the details 
of a cycle.

The derivative of $\theta$ with respect to $t$ is the angular velocity
\begin{equation} \label{a3}
\omega = \frac{d \theta }{d t}.
\end{equation} 
As long as $\omega$ is positive the domino train proceeds. So the question is whether 
$\omega$ remains positive till the foremost collides with the next one.  In order to calculate
$\omega$  we have to understand what happens during a collision and what is  the
equation of motion between collisions. The central function in this calculation is the relation
between the angle $\theta_1$ of domino 1 as function of the angle $\theta$ of the leading domino 0. 

We start the discussion with two dominoes 0 and 1. To ease the notation we replace  from now 
on $s/h$ by $s$ and $d/h$ by $d$.
\begin{figure}[h]
\begin{center}
    \epsfxsize=14cm
    \epsffile{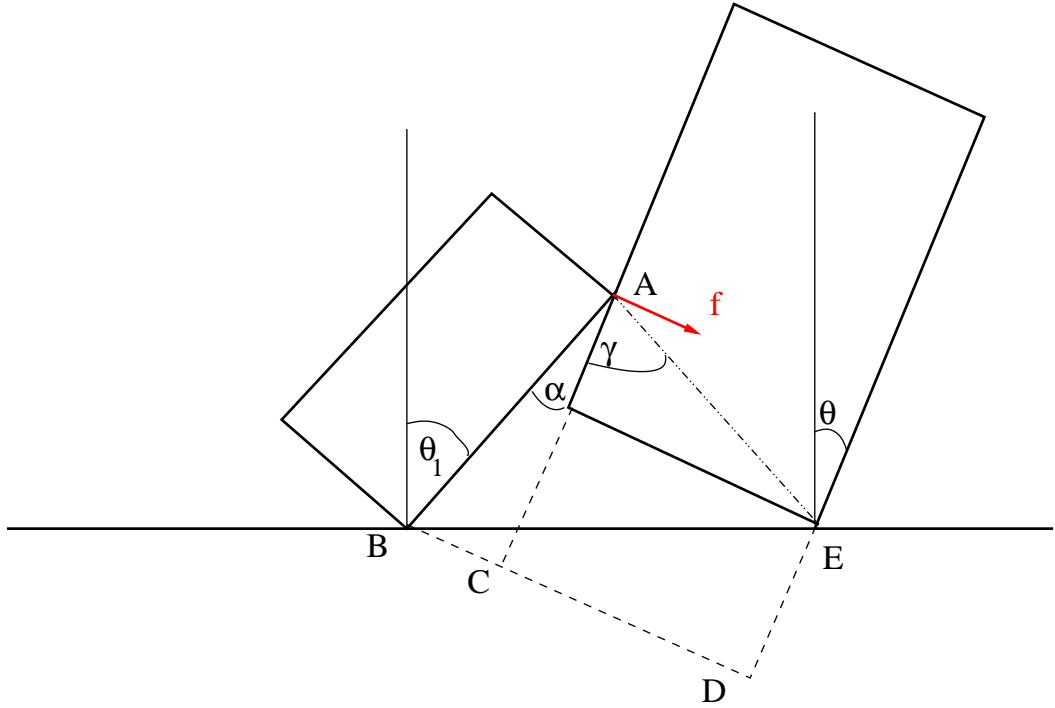}
    \caption{Successive dominoes. The tilt angle $\theta$ is taken with respect to 
the vertical. Domino $1$ hits $0$ at the point A. The rotation axis of $1$ is the point B
and E is that of $0$.
The normal force $f$ that domino 1 exerts on domino 0 is also indicated. The goniometric 
relations are summarized in Table \ref{dominoes}.}  
\label{dominoes}
\end{center}
\end{figure}

\section{Two dominoes} \label{two}

Consider domino 1, freely rotating  towards the still upright domino 0.
In order that domino 1 keeps moving, the initial push must be large enough to overcome 
the point of highest potential energy, occuring at the angle $\theta_u$, given by
\begin{equation} \label{b1}
\sin \theta_u = \frac{d}{\sqrt{1+d^2}}.
\end{equation} 
Here we assume that domino 1 reaches this point before that it hits domino 0. So
we must have $\theta_u < \theta_c$. With (\ref{a2}) and (\ref{b1}) this implies 
\begin{equation} \label{b0}
s > \frac{d}{r \sqrt{1+d^2}}.
\end{equation}
Provided that $s$ fulfils this criterion, we find for the maximum potential energy
\begin{equation} \label{b2}
E_{1u} = \frac{1}{2} \, m_1 g h_1 \, (\cos \theta_{1u} + d \sin \theta_{1u}).
\end{equation}
Under marginal circumstances, domino 1 has a vanishing rotation at this point  and 
thus is its total energy also $E_{1u}$. While falling the potential energy decreases and reaches 
at the collision angle $\theta_c$ the value
\begin{equation} \label{b3}
E_{1c} = \frac{1}{2} m_1 g h_1 (\cos \theta_c + d \sin \theta_c). 
\end{equation} 
The energy difference is converted into kinetic energy. At the collision the kinetic energy 
amounts
\begin{equation} \label{b4}
 \frac{1}{2}\,  I_1 \, \omega^2_{1c}= E_{1u}-E_{1c},
\end{equation} 
where $I_1$ is the moment of inertia of domino 1.
So the angular velocity at the collision is given by (with $\tau_1$ the time scale (\ref{a1})
for domino 1)
\begin{equation} \label{b5}
\tau_1^2 \omega^2_{1c} = (\cos \theta_{1u} + d \sin \theta_{1u})- 
(\cos \theta_c + d \sin \theta_c).
\end{equation} 
We consider the collision as instantaneous. Domino 0 gets an  impuls $F$ from 1, changing
the angular velocity of domino 0 abruptly from 0 to $\omega_0$, given by
\begin{equation} \label{b6}
I \omega_0 = F a,
\end{equation} 
where $a$ is the arm of the torque exerted on domino 0. Through the principle action is
reaction, domino 0 gives an impuls $-F$ to domino 1, such that its 
angular velocity changes from $\omega_{1c}$ instantaneously to $\omega_0$ 
(as we will see in a moment). So we have the equation
\begin{equation} \label{b7}
I_1(\omega_0 - \omega_{1c}) = - Fb,  
\end{equation}
where $b$ is the moment arm of the torque on domino 1 with respect to its turning point. 
For the computation of $a$ and $b$ we have to rely on the goniometry between touching
dominoes which is sketched in Fig. \ref{dominoes}.

\begin{table}[h]
\begin{center}
\begin{tabular}{|l|c|}
\hline
  &  \\*[-2mm]
quantity & formula \\
  &  \\*[-2mm]
\hline
  &  \\*[-2mm]
top angle $\alpha$ of rectangular triangle ABC & $ \alpha=\theta_1-\theta $\\*[4mm]
base BC of triangle ABC & BC $= h \sin \alpha $ \\*[4mm]
top angle $\beta$ of rectangular triangle EBD & $ \beta = \pi /2 - \theta $ \\*[4mm]
base BD of triangle EBD & BD $ = (s+d) \sin \beta $\\*[4mm]
height $h_{cm}$ of center of mass domino $0$ & $2 h_{cm}= h \cos \theta + d \sin \theta $ \\*[4mm]
moment arm $b$ of force $-f$ exerted on 1 & $b = (h/r) \cos \alpha $ \\*[4mm]
moment arm $a$ of force $f$ exerted  & $a = \, $AE$ \, \sin \gamma =\, $AC-DE\\*[4mm]
domino $0$          &  $ = (h/r) \cos \alpha - (s+d) \sin \theta $ \\*[4mm]
\hline
\end{tabular} 
\end{center}
\caption{Goniometric relations referring to Fig.~\ref{gonio}}\label{gonio}
\end{table}  

The various goniometric relations are explained in Table \ref{gonio}.
The basic relation between the angle $\theta_1$ and $\theta$ follows by equating the 
given expression for BC and for BC as BD$-d$
\begin{equation} \label{b8}
(1/r) \sin(\theta_1 - \theta) = (s+d) \cos \theta - d.
\end{equation} 
It is a direct consequence of the condition that the dominoes lean on each other
after a collision.  (\ref{b8}) defines $\theta_1$ as a function $\theta_1(\theta)$ of $\theta$ 
in the interval $0 \leq \theta \leq \theta_c$.

The moment arms $a$ and $b$, as defined in Table \ref{gonio}, are obviously equal at
collision where $\theta=0$. The angular velocity of domino 1 just after collision, equals that
of domino 0. To see this we have to determine the relation between
$\omega_1$ and $\omega$, when the motion of domino 1 is a slave of domino 0. 
Generally holds
\begin{equation} \label{b9}
\omega_1 = \frac{d \theta_1}{d t} = \frac{d \theta_1}{d \theta} \, \omega = 
\theta'_1 (\theta)  \, \omega .
\end{equation} 
The prime denotes differentiation with respect to the argument and is used for shortness. 
Differentiation of (\ref{b8}) with respect to $\theta$ gives
\begin{equation} \label{b10}
\theta'_1 (\theta) =  \frac{(1/r) \cos\, (\theta_1 - \theta)-
(s+d) \sin \theta}{(1/r) \cos\, (\theta_1- \theta)} = \frac{a}{b}.
\end{equation} 
The last equality is demonstrated in Table \ref{gonio}. Using this relation for $\theta=0$,
where $a=b$, leads to $\theta'_1 (0) =1$, showing that the two angular velocities are equal
at collision.

We use arguments for the angle $\theta$ and subscripts for corresponding time. So 
$\omega_{1c}$,  in (\ref{b5}), is the angular velocity of domino 1 just before the collision 
with 0 and $\omega_1 (0)$ is its value just after the collision has taken place. 

The equality $a=b$ (at collision) makes it easy to eliminate the impuls $F$, yielding
the following relation between $\omega_0$ and $\omega_{1c}$
\begin{equation} \label{b11}
(I + I_1) \, \omega_0 = I_1 \,\omega_{1c}.
\end{equation} 
Here we see that the ratio of the moments of inertia of the two dominoes is the important 
ingredient for the collision. It equals $r^{-q-1}$ with $q=4$ for massive dominoes and 
$q=3$ for hollow dominoes.

Now $\omega_0$ must be large enough to get the combination of the two sliding 
dominoes over the maximum of their combined potential energy,
which equals
\begin{equation} \label{b12}
V (\theta) = \frac{1}{2} m g h \, [r^{-q } (cos \theta_1 + d \sin \theta_1) + 
\cos \theta + d \sin \theta].
\end{equation} 
Note that domino 1 weighs by a factor $r^{-q}$ less in the sum. We get a condition for
the domino effect by requiring that the kinetic energy of the pair 1 and 0 must be larger 
than the potential barrier. With $\omega_1(0)=\omega_0$ the condition reads
\begin{equation} \label{b13}
\frac{1}{2} (I + I_1) \, \omega^2_0 \geq [V(\theta_m) - V(0)],
\end{equation} 
where $\theta_m$ is the tilt angle where the maximum of $P(\theta)$ is reached. It will be
close the angle $\theta_u$ where domino 0 reaches its maximum potential energy, but
the contribution of domino 1 is decreasing at this point and so $\theta_m$ is (slightly)
smaller than $\theta_u$ (as the contribution of domino 1 only counts by a factor $r^{-q}$
less). 

We now have sufficient equations to determine maximum factor $r$ still allowing the 
domino 1 to topple over domino 2: the expression (\ref{b5}) for $\omega_{1c}$, 
the relation (\ref{b11}) between $\omega_0$ and $\omega_{1c}$ and the condition 
(\ref{b13}) for $\omega_0$. $\tau_1$ is related to $\tau$ by
\begin{equation} \label{b14}
\tau^2_1=\tau^2 /r.
\end{equation} 
\begin{figure}[h]
\begin{center}

    \epsfxsize=12cm
    \epsffile{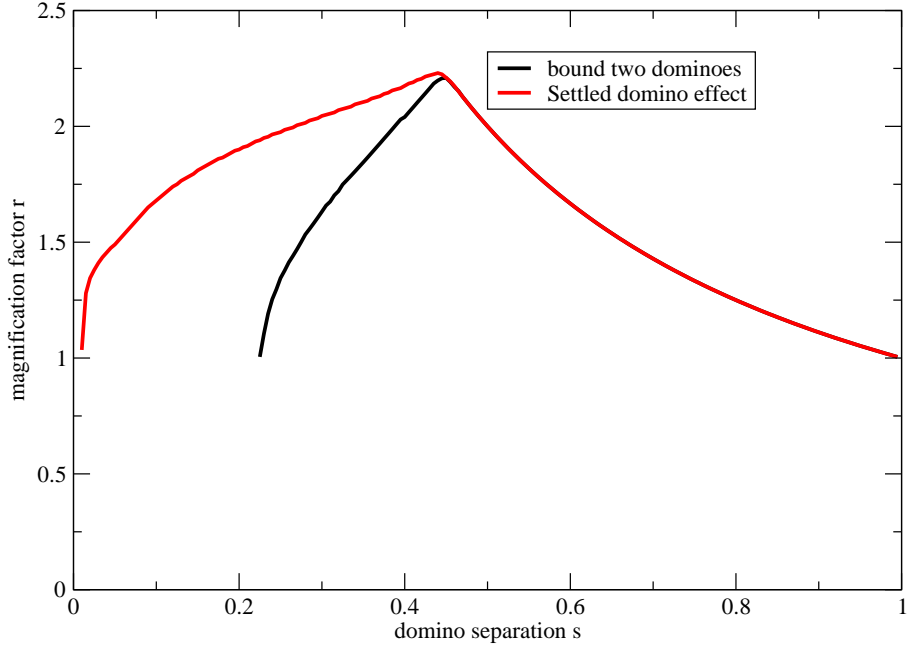}
    \vspace*{4mm}

    \caption{Bounds on the marginal domino effect for hollow dominoes.}  
\label{marginal}
\end{center}
\end{figure}

The above equations impose on the parameter $r$ a condition for given $s$. The boundary,
for which the domino effect becomes marginal, is given by the equality in (\ref{b13}). 
Although the equations can written out explicitly, their solution is too complicated to discuss
analytically, because the angle $\theta_m$ cannot be found analytically. However it is simple
to do it numerically. Start with a value of the separation $s$ (well beyond the restriction (\ref{b0}))
and a low value for $r$ for which we are sure that the condition (\ref{b13}) is fulfilled. 
Then determine $\theta_m$ by slowly raising $\theta$ from $\theta=0$ and see whether 
the maximum $V (\theta)$ is reached. Once the $V(\theta)$ starts to decrease, one is beyond 
the maximum. It may happen right away at $\theta=0$ and then the maximum is $\theta_m=0$,
where (\ref{b13}) is surely fulfilled. Otherwise check whether (\ref{b13}) is fulfilled and if so
raise $r$. If not, one is beyond the boundary, which then can be determined more accurately
by interpolating beteen the present $r$ and the previous $r$.
The outcome of this exercise is shown for hollow dominoes in Fig. \ref{marginal}.
The flank of the curve at the large-separation side is trivial: the domino must be large enough 
to hit the next domino.  

\section{A general domino train} \label{general}

The discussion of the previous section suffers from the fact that we have assumed that 
the initial push to domino 1 is marginal, i.e. just sufficient to get it over its own potential
barrier. The bound following from this condition is too restrictive in practice. We now 
assume that the initial push is large enough to set the train into motion and ask the question
whether the train can sustain itself, i.e. whether the total kinetic energy after collision with
a new domino, is large enough to get the enlarged collection over its potential barrier. 
In every cycle the potential barrier can only shift to smaller angles $\theta$,  since 
at the tail, one more domino is in the train, which is already (long) over its potential maximum. 
Once the maximum occurs at $\theta=0$, there is no question anymore whether the new
train will make it, since it runs only downhill. 

The analysis of this problems runs to a large extent along the same lines as in the 
previous section. We divide out the mass and are concerned about the total height 
contribution to the potential energy of a train of $N$ dominoes leaning on each other. 
It follows as the sum
\begin{equation} \label{c1}
H(\theta) = \sum_{i=0}^{N-1} r^{-q i}  \, [\cos \theta_i (\theta) + d \sin \theta_i (\theta)].
\end{equation} 
Likewise the total kinetic energy can be represented by
\begin{equation} \label{c2}
J (\theta) = \sum^{N-1}_{i=0} r^{-(q+1) i} \, [\theta'_i (\theta)]^2. 
\end{equation} 
Here $\theta_i(\theta)$ is the tilt angle of domino $i$ as function of the tilt angle
$\theta$ of the foremost and $\theta'_i (\theta)$ is its derivative.
In appendix \ref{rotation} we derive that conservation of energy between collisions
implies relation (\ref{d9}) or
\begin{equation} \label{c3}
J(\theta) \, \tau ^2\omega^2 (\theta) =H(0) - H(\theta) + J(0) \, \tau ^2\omega^2 (0). 
\end{equation} 
This enables to calculate $\omega (\theta)$ for given $\omega (0)$. 

(\ref{c3}) also yields
the initial value $\omega(0)$ if we combine it with the collision equation and the 
scale invariance. 
In appendix \ref{coltrain} we show that the angular velocity $\omega_0$ of the
foremost domino, just after being hit, is related to the angular velocity $\omega_{1c}$
of the hitting domino as  
\begin{equation} \label{c4}
J(0) \, \omega (0)  = (J (0) -1)  \, \omega_1 (\theta_c).
\end{equation}

So far this is just a generalization from 2 dominoes to $N$ dominoes in the train. 
The new element is that we require that the cycles are self-similar.
\begin{equation} \label{c5}
\tau_1 \, \omega_1 (\theta) =\tau \, \omega (\theta) \quad \quad {\rm or}
\quad \quad  \omega_1 (\theta) = \sqrt{r} \, \omega (\theta) 
\end{equation} 
The three equations (\ref{c3}-\ref{c5}) determine the values of $\omega (0)$. 
\begin{equation} \label{c6}
\tau^2 \, \omega^2 (0) =  \frac{P \,(J(0)-1)}{ J(0) [r^q J(0) -J(0) +1]}, \\*[4mm]
\end{equation} 
Here $P$ is the ``fuel'' of the domino effect: the difference between the potential
energy of an upright domino and a fallen domino.
\begin{equation} \label{c7}
P=P(h,d,s) = H(0)-H(\theta_c)= 1-\cos \theta_f  - d \sin \theta_f =1-x_f - d y_f.
\end{equation}
The value of $\theta_f$ is given in (\ref{A18}) of Appendix \ref{geometry}.
In (\ref{c6}) we have eliminated the value $J(\theta_c)$ using relation (\ref{d8}).
\vspace*{1cm}

\begin{figure}[h]
\begin{center}
    \epsfxsize=12cm
    \epsffile{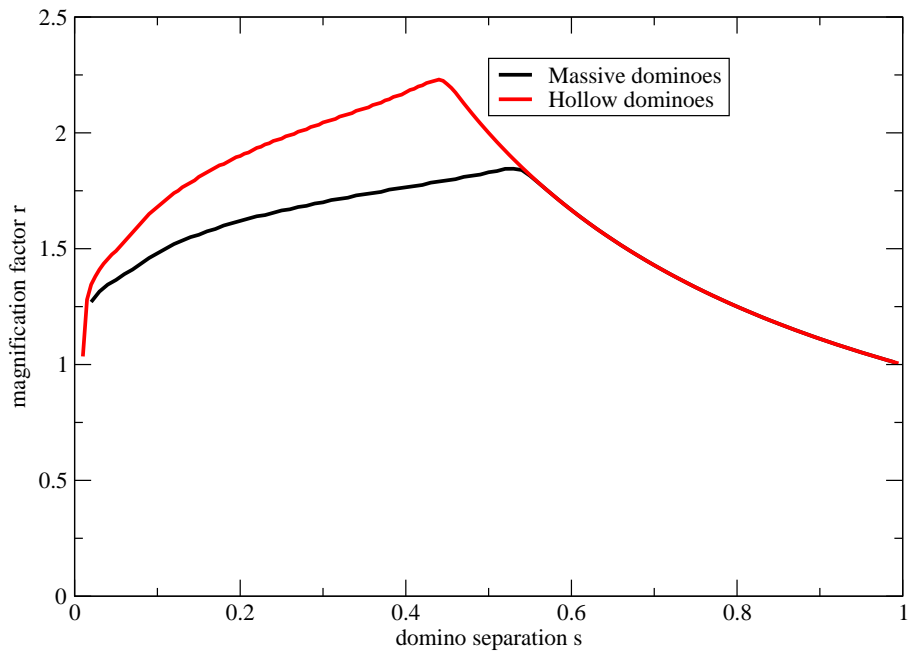}
 
    \caption{Magnification factor as function of the separation.}  
\label{growth}
\end{center}
\end{figure}

With these equations one can check whether the kinetic energy of the train after collision,
exceeds the potential increase from $\theta=0$ to its maximum at $\theta_m$
\begin{equation} \label{c8}
\tau^2 \omega^2_0 \geq [H (\theta_m) - H(0)],
\end{equation} 
We now summarize the steps to find the maximum $r$ with a domino effect as function of $s$.
In the Appendices the derivations and additional  expressions are given.  
\begin{enumerate}
\item Compute for a given $s$ and $r$ the initial value of $x_f$ as given by (\ref{A18})
or (\ref{A16}), given in Appendix \ref{geometry}, depending on the condition (\ref{A17}).
\item Start the iteration scheme for the $\theta_i (\theta)$ and $\theta'_i (\theta)$, as described
in Appendix \ref{geometry}, for the angle $\theta=0$. This yields the value of $J(0)$. 
Compute with (\ref{c7}) the value of the fuel $P$ and with  (\ref{c6}) the initial angular
velocity of a cycle $\omega(0)$.
\item Then search with equation (\ref{c1}) for the value $\theta_m$ for which  
$H(\theta)$ is maximal and find the corresponding minimal $\omega(\theta_m)$ with (\ref{c3}).
\item If $\omega(\theta_m)$ is still positive, raise the value of $r$ and repeat the previous steps 
2-4 in order to find again the minimal angular velocity $\omega(\theta_m)$. Keep raising $r$ 
till the minimum becomes zero. This gives the maximum magnification factor for the 
chosen value of $s$ 
\item Do this story for $0 < s < 1$
\end{enumerate}
The result of the calculation is shown in Fig. \ref{growth} for hollow and massive dominoes.
The curve consists of a rising part
for $s< 0.5$ an a drop-off as $r=1/s$ for higher values. The latter part of the
curve reflects the condition the foremost domino must be hit by the previous in order to
be toppled. This limit is fairly unrealistic, but follows from our condition that the dominoes
cannot slip over the ground but only topple. Near that line, the foremost domino hits the next
at a low point, such that the tendency to slip is larger than to topple, as the moment arm becomes
very small.

\appendix 
\section{Relations between the falling dominoes}\label{geometry}

This Appendix discusses the geometric relations between the row of falling dominoes 
that lean on each other. In Section \ref{two} we gave in equation (\ref{b8}) the relation 
between the tilt angle $\theta_1$ of domino 1 and the $\theta$ of domino 0.
The successively higher numbered dominoes follow from the same relation as domino $i$ 
is related in exactly the same way to $i-1$ as 1 to 0. In general
\begin{equation} \label{A2}
\theta_i (\theta) = \theta_1 (\theta_{i-1}) = \theta_1(\theta_1(\theta_{i-2}) = 
\theta_1(\theta_1(\theta_1 (\cdots (\theta )))) 
\end{equation} 
Thus equation (\ref{A2}) gives all the tilt angles of the followers in terms of
$\theta$. Rather then using the explicit relation 
\begin{equation} \label{A3}
\theta_i (\theta) = \theta_{i-1} (\theta) + \arcsin ( (r/h)[(s+d) \cos \theta_{i-1}(\theta) - d]),
\end{equation} 
we construct the relation in terms of a rotation in carthesian coordinates
\begin{equation} \label{A4}
x_i=\sin \theta_i, \quad  y_i= \cos \theta_i. 
\end{equation} 
With the definitions
\begin{equation} \label{A5}
X_i= \sin(\theta_{i+1}- \theta_i) = (r/h) [(s+d)\, y_i -d], 
\quad Y_i = \cos (\theta_{i+1} -\theta_i)= \sqrt{1 - X^2_i},
\end{equation} 
the rotation equations read (using $\theta_{i+1} = \theta_i + (\theta_{i+1} - \theta_i)$)
\begin{equation} \label{A6}
x_{i+1} = Y_i \, x_i + X_i \, y_i, \quad \quad y_{i+1} = - X_i \, x_i + Y_i \, x_i.
\end{equation} 
These equations are the same for every pair of successive dominoes.
Note that this scheme contains only multiplications and one square root and not any
goniometric function, which speeds up the iteration. 

The function $\theta_1 (\theta)$ is the central ingredient for the calculations. So we
list a few more of its properties. As $\theta=0$ corresponds to the collision angle 
$\theta_c$ for domino 1 we have
\begin{equation} \label{A7}
\theta_1(0) = \theta_c.
\end{equation} 
By differentiating (\ref{A2}) with respect to $\theta$, we  
construct a recursion relation between the derivatives 
\begin{equation} \label{A9}
\theta'_{i+1} (\theta)= \frac{d \theta_1(\theta_i)}{d \theta_i } \, \theta'_i (\theta).
\end{equation} 
For the first factor we can use (\ref{b10}) in the form
\begin{equation} \label{A10}
\frac{d \theta_1(\theta_i)}{d \theta_i } =\frac{(1/r) \, Y_i -(s+d) \,x_i} {(1/r) Y_i}
=\frac{a_i}{b_i},
\end{equation} 
with the convention the indices 0 apply to the foremost. We cast the combination of (\ref{A10}) 
and (\ref{A9}) in the convenient recursive form
\begin{equation} \label{A11}
\theta'_{i+1} \, b_i = \theta'_i \, a_i.
\end{equation} 
As last useful formula we write (\ref{A2}) as
\begin{equation} \label{A12}
\theta_i (\theta) =\theta_{i-1} (\theta_1 (\theta))
\end{equation} 
and differentiate with respect to $\theta$, yielding for the derivative
\begin{equation} \label{A13}
 \theta'_i (\theta) = \theta'_{i-1} (\theta_1(\theta)) \, \theta'_1 (\theta).
\end{equation} 
We will use this expression for $\theta=0$ with the result, using (\ref{A7}) and (\ref{b10}) 
\begin{equation} \label{A14}
\theta'_i (0) = \theta'_{i-1} (\theta_c).
\end{equation} 

The derivatives of the $\theta_i$ are employed in the expression of the angular velocities 
$\omega_i$ in terms of $\omega$
\begin{equation} \label{A15}
\omega_i = \frac{d \theta_i}{ d \theta} \, {d \theta \over dt} = \theta'_i \, \omega.
\end{equation} 

This completes the discussion of the properties of the falling dominoes in terms of
the parameters $\theta$ and $\omega$ of the foremost. The transformations (\ref{A6})
generate for a given $\theta$ the sequence of $\theta_i$ or rather the sets $x_i,y_i$
and $X_i,Y_i$ and with these values we can iterate simultaneously the angular velocities
$\omega_i$ in terms of $\omega$. In principle we would have to iterate $\infty$ many
steps, but the tilt angles rapidly approach the stacking angle $\theta_\infty$,  following as 
\begin{equation} \label{A16}
y_\infty= \cos \theta_\infty = d/(s+d).
\end{equation}
The stacking angle is the angle in which a static row of dominoes leans on each other for 
a given separation $s$. So a limited number of iteration steps suffices. In fact for 
rather wide separation the stacking angle is not realized, because the previous domino is too
small to lean on the next.  This happens when
\begin{equation} \label{A17}
(1/r)^2 < (s+d)^2 - d^2.
\end{equation} 
The last tilt angle $\theta_f$ which enables leaning is given by 
\begin{equation} \label{A18}
x_f= \sin \theta_f = \frac{(1/r)^2+(s+d)^2-d^2}{2 (1/r) (s+d)}.
\end{equation} 
As soon as $x_i$ exceeds $x_f$, one must end the iteration, because for larger angles
the domino does not lean anymore, but falls free and does not contribute 
to the mechanics of the domino effect. In case condition (\ref{A17}) is fulfilled, we take
$x_\infty =x_f$.

\section{The motion between collisions} \label{rotation}

The domino effect is a succession between tumbling and colliding. In this Appendix
we consider the phase of tumbling. As the dominoes slide frictionless over each other,
the motion is completely determined by the conservation of energy. Energy is lost in
the collissions. The energy has a kinetic and a potential part. The potential energy
of domino $i$ is given by 
\begin{equation} \label{d1}
V_i = \frac{1}{2}  m g \, (h \, y_i+d \, x_i )\, r^{-qi}
\end{equation} 
The power of $r^{-qi}$ comes in the expression because the mass goes down
with a factor $r^{1-q}$ and the size with $r^{-1}$ in every step backwards down the train.
The kinetic energy of domino $I$ reads
\begin{equation} \label{d2}
K_i = \frac{1}{2} I_i \omega^2_i
\end{equation}
with the moment of inertia $I_i$
\begin{equation} \label{d3}
I_i = m_i  \, [ h_i^2 + d_i ^2] /3= m \, [h^2  + d^2 ] \, r^{-(q+1)i}/ 3= I  \, r^{-(q+1)i}. 
\end{equation}
The $q+1$ power of the magnification factor follows from the mass and the size squared. 
The total energy of the falling train then equals
\begin{equation} \label{d4}
E = \sum_i (K_i +V_i). 
\end{equation} 
We make the energy dimensionless by dividing it by $m g h/2$ yielding
\begin{equation} \label{d5}
\epsilon = \frac{2 E}{m g h} = H(\theta)  + \tau^2 \, J (\theta) \, \omega^2,
\end{equation} 
with the total dimensionless potential energy $H$
\begin{equation} \label{d6}
H(\theta) = \sum_i   [x_i + (d/h) y_i]\, r^{-qi}\equiv H(\theta)
\end{equation} 
and the total dimensionless moment of inertia is 
\begin{equation} \label{d7} 
J(\theta) = \sum_i r^{-(q+1)i} \,  [\theta'_i (\theta)]^2 \equiv J(\theta).
\end{equation} 
For a given value of $\theta$ all the ingredients for the functions $H(\theta)$ and
$J(\theta)$  are well defined by the recursion relations of the previous section. 
We need in Section \ref{general} the values of $J(\theta)$ for the angles $\theta=0$ 
and $\theta=\theta_c$. Using (\ref{A14}) one finds the relation
\begin{equation} \label{d8}
J(0) = 1 + r^{-(q+1)} \quad \quad {\rm or} \quad \quad  J(\theta_c) = r^{q+1} [J(0) -1].
\end{equation} 

Conservation of energy implies that $\epsilon$ is independent of time of the tilt
angle, such that we have
\begin{equation} \label{d9}
\epsilon = H(0) + J(0) \tau^2 \omega^2 (0)=H(\theta) + J(\theta) \tau^2 \omega^2 (\theta).. 
\end{equation} 
This allows to find $\omega (\theta)$ for a given $\omega (0)$.

\section{The collision equation} \label{coltrain}

The stages of rotational motion are connected by collisions.  In this Appendix we
compute the initial value $\omega (0)$ from the final value $\omega_ (\theta_c)$ 
of the previous cycle. As we mentioned in the introduction, we assume that the two
colliding dominoes stick together after the collision. That means that in the in the
center of mass system the relative kinetic energy is completed dissipated. In other
words the collision is fully inelastic.  We assume that we have a fully developed 
domino train such that the cycles are the similar (upon a factor $\sqrt{r}$): 
before the collision the hitting domino 
has an angular velocity $\omega_1(\theta_c)=\sqrt{r} \omega(\theta_c)$ 
and the upright domino gets a velocity $\omega (0)$ after the collision.

The idea is that during the collision, forces are exerted in a very short time span, such
that the angles do not change during the collision. Instead the angular velocities make
a jump. When domino $1$ hits $0$, its own angular velocity is suddenly reduced
and that of $0$ jumps to the non-zero value $\omega(0)$. The jumps in 
the angular velocity propagate downwards in magnitude, in order  to keep the dominoes
in contact. Therefore the impulses have to propagate downwards in order to realize 
these jumps. The jump of the
foremost domino is from $\omega=0$ before the collision to $\omega(0)$ after the collision
\begin{equation} \label{e1}
I \Delta \omega_0 = I \omega(0) = F_0 a_0
\end{equation} 
The second equality gives the integrated torque during the collision, which is the impuls
$F_0 $ exerted by domino 1 on domino 0 times the arm $a_0$ with respect to the rotation
axis of 0. Likewise the domino 1 feels the impuls $-F_0$ from domino 0 and $F_1$ from
domino 2.
\begin{equation} \label{e2}
I_1 \Delta \omega_1 = I_1 \, [\theta'_1 (0)\, \omega (0) - \theta'_0 (\theta_c) \,\omega_1
(\theta_c)] = F_1 a_1 - F_0 b_0.
\end{equation} 
The first term gives the value of $\omega_1$ as calculated from the foremost domino 0. The
second term is the angular velocity of 1 before the collision when domino 1 was the foremost
with angle $\theta_c$. The factor $\theta'_0 (\theta_c) =1$ is there for reason of generality and 
the index 1 in $\omega_1 (\theta_c)$ is to indicate that here domino 1 is seen as the foremost.
In general domino $i$ receives 
$-F_{i-1}$ from $i-1$  and $F_i$ from  $i+1$. As general equation we get
\begin{equation} \label{e3}
I_i \Delta \omega_i = I_i \, [\theta'_i (0)\, \omega (0) - \,\theta'_{i-1} (\theta_c) \,
\omega_1 (\theta_c) ]=F_i a_i - F_{i-1} b_{i-1}. 
\end{equation} 
We see that impuls $F_i$ occurs in two equations: for domino $i+1$ with a minus sign and
for domino $i$ with a plus sign. We can eliminate the impulses by multiplying
equation for domino $i$ with the factor $s_i$ and add them up. If $s_i$ has the property
\begin{equation} \label{e4}
s_i a_i = s_{i+1} b_i
\end{equation} 
the impulses drop out of the sum. We take $s_0=1$. Comparing this recursion for
the $s_i$ with the recursion (\ref{A11}) for the derivatives of the tilt angles we may 
identify
\begin{equation} \label{e5}
s_i = \theta'_i (0). 
\end{equation} 
The argument $\theta=0$ occurs in this formula, because the moment arms are taken
in the situation where the foremost domino is hit and thus still has a tilt angle $\theta=0$.

Using (\ref{e5}) the coefficient of $\omega (0)$ becomes
\begin{equation} \label{e6}
\sum_{i =0}  I_i  \,  [\theta'_i (0)]^2 = I \, J(0) .
\end{equation} 
The coefficient of $\omega_1 (\theta_c)$ reads with (\ref{A14})
\begin{equation} \label{e7}
\sum_{i=1} \, \theta'_i (0) I_i\, \theta'_{i-1} (\theta_c) = \sum_{i=1}  I_i \, [\theta'_i (0)]^2 = 
I\, (J(0)-1),
\end{equation} 
such that we obtain the collision law
\begin{equation} \label{e8}
J(0) \, \omega (0)  =  (J(0)-1) \, \omega_1(\theta_c).
\end{equation}

\end{document}